\documentclass[letterpaper, 10 pt, conference]{ieeeconf}  

\IEEEoverridecommandlockouts                              

\usepackage[T1]{fontenc}
\usepackage{graphicx}

\usepackage{amsmath}  
\usepackage{amsfonts}  
\usepackage{bm}   
\usepackage{algorithm}
\usepackage{algorithmic}

\usepackage{setspace}  

\title{\LARGE \bf
Formation Control of Multi-agent System with Local Interaction and Artificial Potential Field
}

\author{Luoyin Zhao$^{1,2}$, Zheping Yan$^{2}$, Yuqing Wang$^{3}$ and Raye Chen-Hua Yeow$^{1}$
\thanks{*This work is supported by Major Research Project on Scientific Instrument Development, National Natural Science Foundation of China (No. 42327901) and China Scholarship Council }
\thanks{$^{1}$Luoyin Zhao and Raye Chen-Hua Yeow are with the Department of Biomedical Engineering, National University of Singapore,
        Singapore, 119276, Singapore
        {\tt\small e1113782@u.nus.edu, rayeow@nus.edu.sg}}%
\thanks{$^{2}$Zheping Yan and Luoyin Zhao are with the College of Intelligent Systems Science and Engineering, 
        Harbin Engineering University, Harbin, 150001, China
        {\tt\small yanzheping@hrbeu.edu.cn}}%
\thanks{$^{3}$Yuqing Wang is with the School of Electrical and Electronic Engineering, Harbin University of Science and Technology,
        Harbin, 150080, China
        {\tt\small wangyq@hrbust.edu.cn}}%
}

\begin{document}
\maketitle
\thispagestyle{empty}
\pagestyle{empty}

\begin{abstract}

A novel local interaction control method (LICM) is proposed in this paper to realize the formation control of multi-agent system (MAS). A local interaction leader follower (LILF) structure is provided by coupling the advantages of information consensus and leader follower frame, the agents can obtain the state information of the leader by interacting with their neighbours, which will reduce the communication overhead of the system and the dependence on a single node of the topology. In addition, the artificial potential field (APF) method is introduced to achieve obstacle avoidance and collision avoidance between agents. Inspired by the stress response of animals, a stress response mechanism-artificial potential field (SRM-APF) is proposed, which will be triggered when the local minimum problem of APF occurs. Ultimately, the simulation experiments of three formation shapes, including triangular formation, square formation and hexagonal formation, validate the effectiveness of the proposed method. 

\end{abstract}

\section{INTRODUCTION}
Formation control is crucial for multi-agent system (MAS) in performing cooperative missions. MASs, which have the characteristics of autonomy and abstract structural units, usually consist of a network of interacting, movable physical entities that can work together to complete complex tasks beyond their individual capabilities. They can be unmanned aerial vehicles \cite{c1},  mobile agents\cite{c2}, autonomous underwater vehicles\cite{c3}, satellites, etc. With the continuous development of high-tech and modern science, MASs have become indispensable in various fields, especially in local wars, resource exploration and disaster relief. However, the effective control of the formation is central to ensuring the safe execution of these missions.

Formation control aims at designing the control protocol to make the MAS form the pre-specified geometry structure and maintain the corresponding formation during the execution process of the task. Usually, when designing a control protocol, the constraints imposed by physical conditions on the agent state variables will also be taken into consideration. Due to the changes in perception and interaction methods, a series of formation control problems are studied.

At present, studies on formation control have made certain progress. Generally, MAS formation control methods include the leader-follower method (LFM)\cite{c5}, virtual structure method (VSM)\cite{c6}, behavior-based control method (BBCM)\cite{c7}, consensus control method (CCM)\cite{c8}\cite{c9}, etc. For the leader-follower method, one or more agents will be chosen as leaders and the rest agents are the followers \cite{c10}. The leader will move along the predetermined trajectory or the real time planning path for the leaders. And the followers will adjust their states to align with the leader by information exchanging while maintaining the spatial constraints of the designated formation. The LFM is one of the most widely used methods in formation control since it is not only easy to design topology and execute, but also facilitates formation scale adjustment\cite{c11}. However, when the number of followers in the system is large, it will be a huge challenge for the communication equipment of the leader. VSM is proposed by Lewis et al\cite{c12}. It regards the MAS as a virtual single rigid body structure, and each agent has a fixed relative position in this structure. The whole system will move with the virtual geometric center, which is convenient to realize the bilateral control. It is often used in satellite formation control, but it has to move according to a certain virtual rigid body structure, its application area is limited. BBCM is introduced by Balch and Arkin in formation control\cite{c13}. The idea is to decompose formation control into a series of basic behaviors, including maintaining the pre-designed formation, moving to the goal, obstacle and collision avoidance et al. Formation  control is achieved through comprehensive weighting of behaviors, which enables each individual in the system to complete tasks in collaboration with other individuals based on their own decisions. However, this method forms control instructions based on preset information and trigger conditions, so it will reduce the flexibility and adaptability of the system. CCM means that the agent uses the state information of the adjacent individuals to update its own state information, thereby making all agents in the system achieve consistent status\cite{c14}\cite{c15}. It is a distributed control structure that has good flexibility and increases the robustness of the system. The system will not have a big impact on the overall formation due to the damage of one individual.

Usually, the MAS works in hazardous environment, which includes uncertain obstacles, so it's important to consider the influence of the obstacles\cite{c16}. Meanwhile, the collision avoidance between agents should also be considered. Artificial potential field (APF) is useful method in obstacle avoidance, however it has the local minimum problem (LMP) where the attraction equals the repulsion, even though it is not the target point\cite{c17}\cite{c18}.

In this paper, we propose a new local interaction control method (LICM), which combines the advantages of the LFM and CCM to reduce the communication burden between agents and keep the flexibility of the formation at the same time. And a stress response mechanism (SRM) is provided to solve the local minimum problem of traditional APF, which is utilized to avoid obstacles and collisions.

\section{PROBLEM AND PRELIMINARIES}

Consider a system with $N$ agents whose initial positions are randomly distributed. And the system runs in an environment with obstacles. The goal is to design the controller so that the system drives to its destination in the desired formation $\mathcal{F}$ while also avoiding obstacles in the environment.

\subsection{Model of Multi-agent System }
For an agent group with $N$ individuals, the dynamics of the system can be represented as:
$$
\dot{\bm{q}_{i}}=\bm{u}_{i}, \quad i=1,2,\cdots,N \eqno{(1)}
$$

Where $\bm{q}_{i}=[x,y]^{T}$ denotes the state vector of the $i$-th agent, $x$, $y$ are position coordinates,  $\bm{u}_{i}$ denotes the control input of $i$-th agent, and $\bm{q}_{i},\, \bm{u}_{i} \in \mathbb{R}^{2}$, $\bm{q},\, \bm{u} \in \mathbb{R}^{2N}$, and $i\in\mathcal{V}$, $\mathcal{V}=\{1,2,\cdots,N\}$. The formation of $N$ agents can be mathematically represented as: $\mathcal{F}=\{q_{1}^{T},q_{2}^{T},\cdots, q_{N}^{T}\}^{T}$, correspondingly the control input can be denoted as: $\bm{u}=\{u_{1}^{T},u_{2}^{T},\cdots, u_{N}^{T}\}^{T}$.

\subsection{Graph Theory}

Graph theory is an efficient tool used to portray the relationship between agents. Usually, a graph can be represented by $\mathcal{G} =\{\mathcal{V}, \mathcal{E}, \mathcal{A}\}$, where the vertex set is $\mathcal{V}=\{1,2,\cdots,N \}$, which denotes the agents in the system. The edge set is $\mathcal{E}=\{(i,j)|i,j \in \mathcal{V},i \neq j \}$, and $\mathcal{E}\subseteq \mathcal{V}\times\mathcal{V}$, which denotes the connection between the agents. $\mathcal{A}= [a_{ij}]$ is the adjacent matrix representing the communication weights, and $a_{ij}\in\{0,1\}$\cite{c19}.

The graph $\mathcal{G}$ is connected if there is path from any node to the other nodes. For undirected graph, the ordered pairs of $(i,j)$ and $(j,i)$ denote the same edge. The max length for any pair of vertices in $N$-node connected graph is $N-1$.

The neighborhood set of the $i$-th agent can be defined as: $   \mathcal{N}_{i}=\{j:j \in \mathcal{V},(i,j)\in \mathcal{E}\}   $. For a distributed formation system, the $i$-th agent only interacts with its neighbor $j$, and $j \in \mathcal{N}_{i}$.

\begin{figure}[thpb]
      \centering
      \includegraphics[scale=1.0]{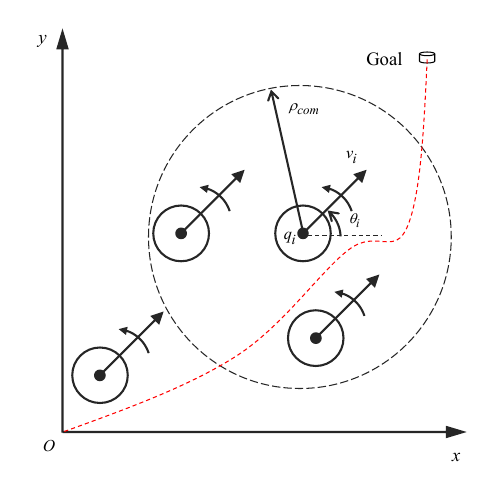}
      \caption{Diagram of the topology for $N$ agents }
      \label{fig:Diagram of the formation topology }
\end{figure}

\subsection{Rigid Formation}

For a multi-agent formation system, if the internal relative distances between agents  remain constant, it is called a rigid formation. The relative position vector between the $i$-th agent and its neighbour $j$ can be represented as: $\bm{r}_{ij}=\bm{q}_{i}-\bm{q}_{j}$, $\bm{q}_{i}=(x_{i},y_{i})$, $\bm{q}_{j}=(x_{j},y_{j})$, and the relative distance between agent $i$ and $j$ is the second norm of their relative position vector $\bm{r}_{ij}$ in Euclidean space, which can be represented as:
$$\Vert \bm{r}_{ij} \Vert=\Vert \bm{q}_{i}-\bm{q}_{j}\Vert=\sqrt{(x_{i}-x_{j})^{2}+(y_{i}-y_{j})^{2}} \eqno(2)$$

As shown in Fig.\ref{fig:Diagram of the formation topology }, the position of the $i$-th agent is $\bm{q}_{i}$, the speed is $v_{i}$, and the direction angle is $\theta_{i}$, equipped with the positioning and communication system, the agent can sense the position information of its  neighbors, the maximum communication distance is $\rho_{com}$, and $\rho_{com}>0$. In two dimensional space, the communication range of the agent is an open ball with the agent as the center and the maximum communication distance $\rho_{com}$ as the radius. Other agents within the communication range of agent $i$ are its neighbors, and the neighbourhood set of agent $i$ can be further constrained as:
$$\mathcal{N}_{i}=\{j:j\in \mathcal{V},\Vert \bm{r}_{ij} \Vert \leq \rho_{com}\}\eqno(3)$$

For a rigid formation $\mathcal{F}$, the distance between the $i$-th and $j$-th agent $\Vert \bm{r}_{ij} \Vert$ is a preassigned constant.

\subsection{SRM-Artificial Potential Field}

\begin{figure}[thpb]
      \centering
      \includegraphics[scale=1.0]{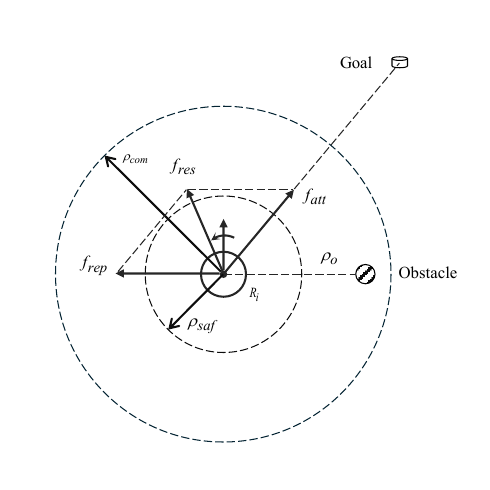}
      \caption{Diagram of obstacle avoidance with artificial potential field }
      \label{fig:obstacle}
\end{figure}

The SRM-Artificial Potential Field (SRM-APF) is introduced in this paper to realize the obstacle avoidance for the multi-agent formation (MAF). As shown in Fig. \ref{fig:obstacle}, in the artificial potential field, the agent is driven towards the goal by the attractive force from the goal point. When it approaches the action scope of the obstacle, a repulsive force will be generated to drive the agent away from the obstacle. Finally, the agent will move to the target under the resultant force\cite{c16}\cite{c20}.

The attractive force is related to the distance from the agent to the target point $\bm{q}_{t}$, and the distance $\rho_{t}$ is represented as:
$$ \rho_{t} = \Vert \bm{q}_{t}-\bm{q}_{i}\Vert \eqno(4)$$

Correspondingly, the attractive force acting on the agent from the target is defined as:
$$ f_{att}(\rho_{t})=-\eta \rho_{t} \eqno(5)$$

Where, $\eta$ is the proportional gain factor. Similarly, the repulsive force from an obstacle is related to the distance from the agent to the obstacle $\bm{q}_{o}$. This distance can be defined as:
$$ \rho_{o} = \Vert \bm{q}_{i}-\bm{q}_{o}\Vert \eqno(6)$$

When the agent enters the action scope of the obstacle, the repulsive force it receives is:
$$
f_{rep}(\rho_{o},\rho_{m})=
\left \{
\begin{aligned}
& k_{r}\left(\frac{1}{\rho_{o}}-\frac{1}{\rho_{m}} \right) \frac{1}{\rho_{o}^{2}}\nabla \rho_{o}, & \rho_{o}\leq \rho_{m}\\
& 0, & \rho_{o}>\rho_{m} 
\end{aligned}
\right.
\eqno(7)
$$

Where, $k_{r}$ is the gain factor, and $\rho_{m}$ is the maximum action distance of the obstacle.
In the artificial potential field, the resultant force on the agent is:
$$f_{res}=f_{att}+f_{rep} \eqno(8)$$

Driven by the resultant force, the multi-agent formation will move to the target. However, the LMP will occur when $f_{res}=0$ and $\rho_{t}=0$. Inspired by the stress response of animals, we regard the LMP as a kind of stimulation from the environment, and a new stress response mechanism (SRM) is proposed to solve the LMP, which is defined as:
$$
f_{res}(f_{att},f_{rep},\delta)=
\left \{
\begin{aligned}
& f_{att}+f_{rep}+ \Gamma,&  LMP \quad occurs\\
& f_{att}+f_{rep}, & \quad  No \quad LMP
\end{aligned}
\right.
\eqno(9)
$$

Where, $\Gamma\in (0,1)$. And $\Gamma$ can be triggered only when LMP occurs, which is similar to the bio process of stress response. The $\Gamma$ will break the balance for LMP, and drives the MAS to move forward continually.

\section{METHOD}

A LICM is proposed with the leader follower and consensus frame (LFCF), which regards the $N$-th agent as the leader and the other agents as the followers, and the agents only interact with their neighbours. And the SRM-APF is utilized to realize the obstacles avoidance and collision avoidance between agents.
\subsection{Control Protocol Design}

Consider a MAS performing the formation task in a two-dimensional space. The state vector of the $i$-th agent is $\bm{q}_{i}$, and $\bm{q}_{i}=(x_{i},y_{i})^{T} \in \mathbb{R}^{2}$, and the control protocol can be designed with consensus as:
$$\bm{u}_{i}=\sum_{j\in \mathcal{N}_{i} } a_{ij}(\bm{q}_{j}-\bm{q}_{i})\eqno(10)$$

Where $a_{ij}\in \mathcal{A}$, $\mathcal{N}_{i}$ is the neighbour set of agent $i$, $\bm{q}_{j}$ denotes the state vector of neighbour. Discretizing (1), we can get:
$$\bm{q}[k+1] =\bm{q}[k]+\bm{u}[k] \eqno(11)$$

Here, it is assumed that the leader leads other agents to complete a task requiring certain distance of movement, starting from the starting point and moving to a target point. The control input of the followers in the formation takes the following form:
$$ \bm{u}_{i}=\epsilon \sum_{j\in\mathcal{N}_{i}} a_{ij} (\bm{q}_{j}-\bm{q}_{i}-\bm{r}_{ij}) \eqno(12)$$

Where $\epsilon >0$, and the control input of the leader is:
$$ \bm{u}_{N}=  \gamma D + \sum_{j\in \mathcal{N}_{j}}a_{Nj}\bm{r}_{Nj} \eqno(13)$$

Where $\bm{r}_{Nj}$ denotes the relative position between the leader and the followers, $\gamma$ is a constant, $D$ is the distance from the leader to target. The speed of the leader is affected by adjacent nodes and the distance from the leader to the target point.

Assume that at time $k$, the agent can scan the obstacle coordinates $q_{obs}$ within a certain range obstacle (including other agents), obstacles will have a repulsive effect on the speed of agent $j$, the repulsive force effect $f_{obs}$ satisfies:
$$f_{j}[k]=\sum_{l=1}^{M}f_{rep}(q_{j}[k],q^{l}_{obs}) \eqno(14)$$

Where $M$ denotes the number of obstacles. Here the repulsive act will have an influence on the velocity control of agent, and the control input of the leader for this situation will be: 
$$ \bm{u}_{N}[k]= \Gamma + \gamma D + \sum_{j\in \mathcal{N}_{j}}a_{Nj}\bm{r}_{Nj}+ \mu \cdot f_{N}[k] \eqno(15)$$

And $\mu$ is a constant, and the control input for the followers is:
$$ \bm{u}_{i}[k]=\epsilon \sum_{j\in\mathcal{N}_{i}} a_{ij} (\bm{q}_{j}-\bm{q}_{i}-\bm{r}_{ij})+\mu \cdot f_{i}[k]\eqno(16)$$

And the influence of repulsive force enables the agent to effectively avoid obstacles and collision while maintaining the original formation to the maximum extent, and the repulsive force will become larger as the distance becomes closer, thereby ensuring the safety of the agent.

\subsection{Stabilization Analysis}

The Lyapunov function $V$ is designed as a function related to potential energy. If it can be proven that the error is asymptotically stable over time, the system can be proven to be asymptotically stable.

$$V=\frac{1}{4}\sum_{(i,j)\in \Tilde{E}}\delta_{ij}^{T}\delta_{ij}\eqno(17)$$
$$\dot{V}=\frac{1}{2}\sum_{(i,j)\in \Tilde{E}}\delta_{ij}^{T}\dot{\delta_{ij}} \eqno(18)$$

The relative position between agents $i$ and $j$ is:
$$\bm{r}_{ij} = \bm{q}_{i}-\bm{q}_{j} \eqno(19)$$

Let the error between the actual distance and the expected distance of the $i$-th and $j$-th agents be:
$$\omega_{ij}=||\bm{q}_{i}-\bm{q}_{j}||-\Tilde{d_{ij}} \eqno(20)$$

Here, the error variable is defined as:
$$\delta_{ij}=q_{ij}^{2}-d_{ij}^{2} \eqno(21)$$

 Then,
$$
\begin{aligned}
       \dot{V}&= \frac{1}{2}\sum_{(i,j)\in E}\delta_{ij}^T\dot{\delta}_{ij} \\
& =\frac{1}{2}\sum_{(i,j)\in E}(\omega_{ij}^{2}+2\omega_{ij}\Tilde{d}_{ij})(2\omega_{ij}\dot{\omega}_{ij}+2\Tilde{d}_{ij}\dot{\omega}_{ij})\\
& = \sum_{(i,j)\in E}\omega_{ij}(\omega_{ij}+2\Tilde{d}_{ij})\frac{x_{ij}^{T}(\bm{u}_i-\bm{u}_j)}{\omega_{ij}+\Tilde{d}_{ij}}(\omega_{ij}+\Tilde{d}_{ij})\\
& = \sum_{(i,j)\in E}\omega_{ij}(\omega_{ij}+2\Tilde{d}_{ij})\cdot d_{ij}^{T}\cdot (\bm{u}_i-\bm{u}_j)\\
& = \bm{\delta}^{T} R_{G}(x)\bm{u}\\
& = \bm{\delta}^{T} R_{G}(x)(-\beta R_{G}^{T}(x) \bm{\delta}) \\
& = -\beta \bm{\delta}^{T} R_{G}(x)R_{G}^{T}(x) \bm{\delta}
\end{aligned}
\eqno(22)
$$

Where $R_{G}(x)$ is the rigidity matrix, since the frame $\mathcal{F}$ has infinitesimal rigidity, then the rigidity matrix $R_{G}(x)$ has full rank, and $R_{G}(x)R_{G}^{T}(x)$ is positive definite, we can get:

$$
\begin{aligned}
       \dot{V} & \leq  -\beta \lambda_{min} (R_{G}(x)R_{G}^{T}(x) )\bm{\delta}^{T}\bm{\delta}\\
        & = -\beta\lambda_{min}  (R_{G}(x)R_{G}^{T}(x))V
\end{aligned}
\eqno(23)
$$

According to the derivation, it can be known that the derivative of the energy error function $\dot{V} < 0$, that is, $V$ decreases monotonically with time. In addition, because $\beta$ is a positive number greater than 0, $V$ is always greater than 0. Therefore, when $t\rightarrow \infty $, $V \rightarrow 0$ . It can be seen that the Lyapunov function $V$ decreases monotonically and the system tends to be stable.



\section{ALGORITHM DESIGN}

For a multi-agent system, the process of agents performing tasks collaboratively can be divided into three steps. First, the agent formation shape must be gradually formed after being launched from the starting position. Secondly, the whole formation will drive to the target position as the task requirement. In addition, obstacle avoidance must be performed in order to maintain a safe journey.

The whole process can be realized with the local leader follower frame with SRM-APF method provided in this paper, which is presented in algorithm 1. To begin with, the coordinates of the $N$ agents and the obstacles in the environment will be configured, and the topology of the formation should be specified, and other hyper-parameters should also be assigned. Then, the row 1-14 is the main loop. The leader will go forward to the target under the act of the resultant force $f_{res}$ calculated in line 2 according to (8), and the followers will follow the leader in terms of the pre-specified geometry structure. SRM-APF is used to avoid obstacles and collisions between agents and obstacles, as well as between agents. If the agent meets the local minimum problem, the SRM will be triggered to enable the agent escape from there. (16) in line 8 is used to calculate the $u_{i}$. $k$ is the iteration counter, which will be updated in line 13.

\begin{algorithm}[htbp]
\renewcommand{\algorithmicrequire}{\textbf{Input:}} 
\renewcommand{\algorithmicensure}{\textbf{Output:}}  

%
%
\begin{algorithmic}[1]    
\label{Al:Consensus with Obstacle Avoidance}
\setstretch{1.25}
\REQUIRE{Initialize the states of all agents $\bm{q}_{ini}$, obstacle positions $\bm{q}_{obs}$, target location $\bm{q}_{t}$, the desired topology $\mathcal{A}$ and other hyper-parameters. 
}
\ENSURE{The states of all the agents along the whole path $\bm{q}_{sta}$}
\FOR{$k<k_{max}$}
    \STATE Calculate the attractive force $f_{att} $, the repulsive force $f_{rep} $ and the resultant force $f_{res}$\\
    \IF{stuck in local minimum}
        \STATE $f_{res} \gets \Gamma $ \COMMENT{SRM}
    \ENDIF\\
    \FOR{$i \leq N$}
        \IF{$j \in \mathcal{N}_{i}$}
             \STATE Compute $u_{i}[k]$\\
             \STATE UpdateStates($q_{i}[k]$,$u_{i}[k]$) $\rightarrow$ $q_{i}[k+1]$
        \ENDIF \\

    \ENDFOR
    \STATE CheckIfArrived ($\bm{q}_{Leader},\bm{q}_{Target}$)\\
    \STATE $k+1 \rightarrow k$
\ENDFOR
    \RETURN {$\bm{q}_{sta}$}
\end{algorithmic}
\caption{Local Interaction Formation Control} 
\end{algorithm}

\section{SIMULATION}
In this paper, several simulation experiments are conducted to test the local interaction leader follower frame with SRM-APF MAS formation control method. The initial position coordinates and directions of the agent are randomly configured. And the obstacles in the environment is $\bm{q}_{obs}=[0,1.5;4,3;3,8.8;7,5;15,16]$, $k_{max}=800$, $\rho_{m}=1$, $\bm{q}_{t}=(14,14)$. Experiment 1 to 3 are tested in the same obstacle environment, and we assume that two-way communication can be carried out between agents, that is, $\mathcal{G}$ is an undirected network, and the communication condition is normal.

\begin{figure}[htbp]
      \centering
      \includegraphics[scale=1]{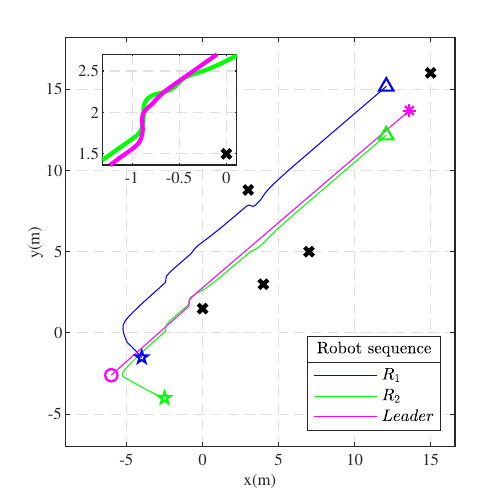}
      \caption{The path for triangle formation in 2D space }
      \label{fig:TrianglePath}
\end{figure}

\subsection{Experiment 1: Triangle formation shape}

The triangle formation is a common formation type, there are three agents in triangular formation, that is, $N=3$. The initial coordinates and direction are randomly pre-specified as $\bm{q}_{ini}=[-4, -1.5, 0;-2.5, -4, pi/4;-6, -2.6, -pi/4]$, the adjacency matrix is $\mathcal{A}=[0,1,1;1,0,1;1,1,0]$, and the relative position error is $\Tilde{\bm{d}}=[-1.5,-1.5,0;1.5,-1.5,0]$. As can be seen from Fig. \ref{fig:TrianglePath}, the curves depict the path changes of the agents, the black x symbols in the picture represent the location of the obstacles. The agents start from the initiate point and gradually form a triangular formation with LICM, and the leader will be driven by $f_{att}$. The formation will drive straight to the target if there are no obstacles in the environment. However, some cured parts appear on the curves when the formation approaches the obstacles. It means that the agent enters the action scope of the obstacles, under the action of SRM-APF, the agent moves away from the obstacle. The local enlarged figure shows the trajectory when the formation approaches the first obstacle. It can be clearly seen that the trajectory is curved, which shows that the SRM-APF really works to realize the obstacle avoidance.
\begin{figure}[htbp]
      \centering
      \includegraphics[scale=1]{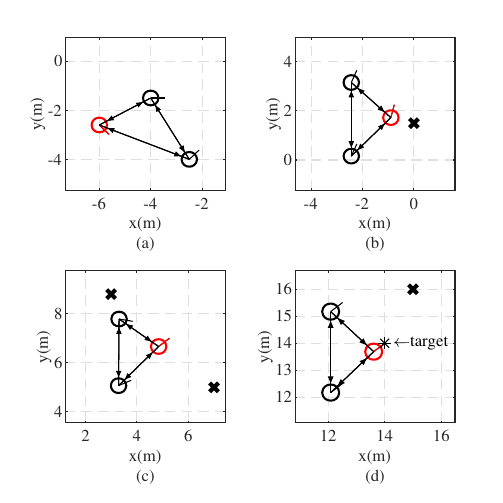}
      \caption{Triangle formation in 2D space }
      \label{fig:TriangleformationScenes}
\end{figure}

 \begin{figure}[htbp]
      \centering
      \includegraphics[scale=1]{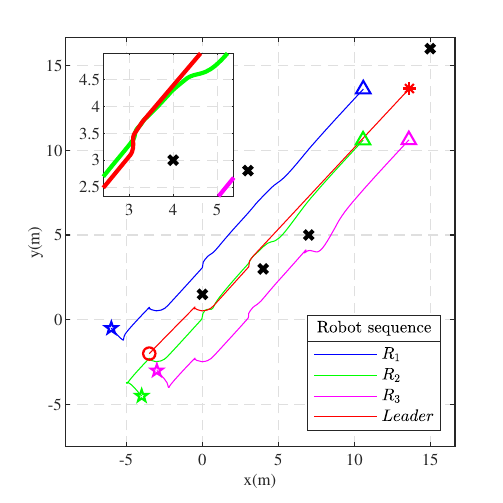}
      \caption{The path for square formation in 2D space }
      \label{fig:SquarePath}
\end{figure}

 Furthermore, Fig. \ref{fig:TriangleformationScenes}  shows four scenes during the formation driving process. Fig. \ref{fig:TriangleformationScenes}(a) shows the random  initial state of the agents. Fig. \ref{fig:TriangleformationScenes}(b) is the scene where the leader encounters the first obstacle. The moving direction of the leader changes under the action of SRM-APF. Fig. \ref{fig:TriangleformationScenes}(c) shows that the direction of the follower changes when approaching the action scope of the obstacle. The formation finally reaches the target point, which is shown in Fig. \ref{fig:TriangleformationScenes}(d). The four scenes show that the formation can maintain the designated formation to reach the target and avoid obstacles at the same time.

\subsection{Experiment 2: Square formation shape}

\begin{figure}[htbp]
      \centering
      \includegraphics[scale=1]{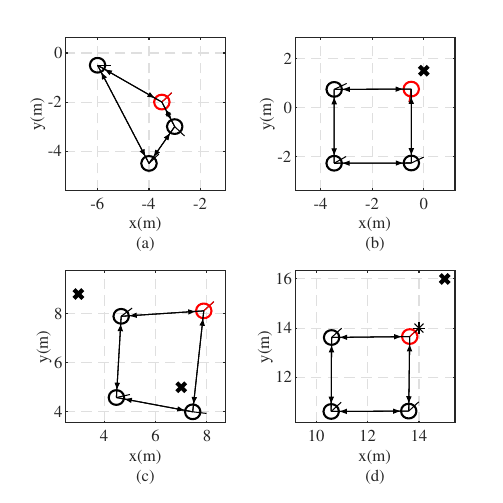}
      \caption{Square formation in 2D space }
      \label{fig:SquareFormationScene}
\end{figure}

For square formation, there are four agents, $N=4$. The states are randomly pre-specified as $\bm{q}_{ini}=[-6, -0.5, 0; -4, -4.5, pi/4;-3, -3, -pi/4;-3.5, -2, pi/4]$, $\mathcal{A}=[0, 1, 0, 1; 1, 0, 1, 0;0, 1, 0, 1;1, 0, 1, 0]$, and the relative position error is $\Tilde{\bm{d}}=[-3, -3, 0,  0;0, -3, -3, 0]$. Fig. \ref{fig:SquarePath} shows the path the formation from the initial positon to the end, the obstacles are labeled with the black x symbols. The four agents quickly formed the square formation under the act of LICM. The curves in the figure reflect the trajectory changes of each agent. The partial enlargement figure shows the impact on the formation of the second obstacle encountered during the formation marching. It can be seen that the repulsive effect is applied to $R_{2}$ and the $Leader$, but has no effect on other agents because they have not traveled to the scope of the second obstacle.

Similar to the triangle formation, four scenarios were tested on the square formation, which is shown in Fig. \ref{fig:SquareFormationScene}. Fig. \ref{fig:SquareFormationScene}(a) shows the initial states for square formation, the agents are randomly distributed. Fig. \ref{fig:SquareFormationScene}(b) and Fig. \ref{fig:SquareFormationScene}(c) show the situation when the leader and follower enter the scope of the obstacle respectively. The direction of the agent that enters the scope of the obstacle first will be the first to change, which will then affect other members of the formation. \ref{fig:SquareFormationScene}(c) indicates that the formation ultimately reaches the target point while maintaining a square formation.   

\subsection{Experiment 3: Hexagon formation shape}

\begin{figure}[h!]
      \centering
      \includegraphics[scale=1]{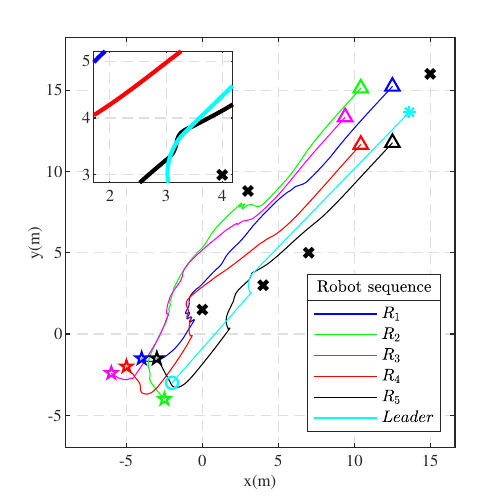}
      \caption{The path for hexagon formation in 2D space }
      \label{fig:HexagonPath}
\end{figure}

There are six agents in hexagon formation, $N=6$, and the $N$-th agent is the leader. The initial coordinates and direction of the agents are randomly pre-specified as:

$$\bm{q}_{ini}=
\begin{gathered}
    \begin{bmatrix} -4& -1.5& 0\\ -2.5& -4 & pi/4 \\-6 & -2.4 & -pi/4\\
 -5& -2& pi/4\\-3 & -1.5 & pi/6 \\  -2& -3& -pi/6
\end{bmatrix}
\end{gathered}
\eqno(24)
$$

And the adjacency matrix is:

$$\mathcal{A}=
\begin{gathered}
    \begin{bmatrix} 0& 1& 0& 0& 0& 1\\ 1& 0& 1& 0& 0& 0\\0& 1& 0& 1& 0& 0\\0& 0& 1& 0& 1& 0\\0& 0&0& 1& 0& 1\\1& 0& 0& 0& 1& 0
\end{bmatrix}
\end{gathered}
\eqno(25)
$$

\begin{figure}[h!]
      \centering
      \includegraphics[scale=1]{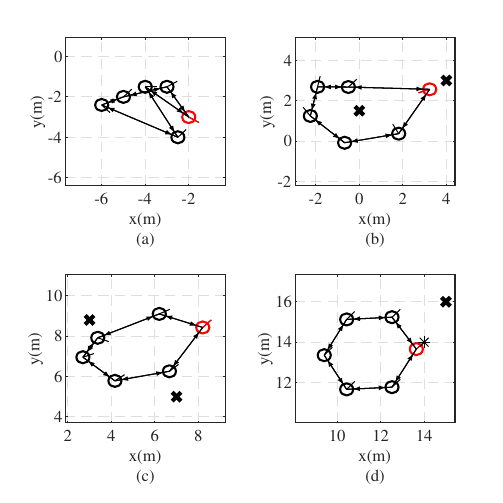}
      \caption{Hexagon formation in 2D space }
      \label{fig:HexagonFormationScene}
\end{figure}

The relative position errors between leader and followers are $\Tilde{\bm{d}}=[-1, -3, -4, -3, -1, 0; \sqrt{3}, \sqrt{3}, 0, -\sqrt{3}, -\sqrt{3}, 0]$. As shown in Fig. \ref{fig:HexagonPath}, the graph shows the trajectories of the agents in hexagon formation from the launching position $\bm{q}_{ini}$ to the target point $\bm{q}_{Target}$. And $\bm{q}_{ini}$ is assigned randomly, the black x symbols in the figure represent the location of the obstacles. Compared with Fig. \ref{fig:TrianglePath} and Fig. \ref{fig:SquarePath}, more curved parts appear in Fig. \ref{fig:HexagonPath}, which means SRM-APF is triggered more times. The curved parts of the partial enlargement graph in Fig. \ref{fig:HexagonPath} show that the obstacles do have an impact on the formation. However, it can still be found that some parts of the curve away from obstacles are also bent. This is because when forming the formation, due to the large number of members in the formation and the close distance between the agents, SRM-APF is triggered to avoid collisions.

Similar to Fig. \ref{fig:TriangleformationScenes} and Fig. \ref{fig:SquareFormationScene}, Fig. \ref{fig:HexagonFormationScene} shows four scenes of the hexagon formation from the starting scene Fig. \ref{fig:HexagonFormationScene}(a) to the target scene Fig. \ref{fig:HexagonFormationScene}(d). Fig. \ref{fig:HexagonFormationScene}(b) and Fig. \ref{fig:HexagonFormationScene}(c) show the impact of obstacles on the leader and follower respectively as the formation advances towards the target. It can be seen that although the obstacles in the environment will affect the formation, the formation can still dynamically adjust the shape to avoid obstacles and collisions, and finally reach the target while maintaining the desired hexagon formation shape.

\section{CONCLUSIONS}

   
In this paper, a novel local interaction formation control scheme with SRM-APF is proposed to realize the formation control of multi-agent system in obstacle environment. The communication overhead of over-reliance on a single point is reduced and the flexibility of the formation is improved by coupling the advantages of leader-follower framework and information consensus. The SRM is proposed to solve the local minumum problem of APF, and the SRM-APF is embbed into the LFCF. Under the action of SRM-APF, the formation has good obstacle avoidance and collision avoidance performance, and it also provides navigation for the formation. The effectiveness of the proposed method has been verified through the simulation experiments, which also means that the proposed method provides a new solution to ensure the safe collaborative operation of multi-agent systems.

\addtolength{\textheight}{-12cm}   





\end{document}